\documentclass[runningheads]{llncs}
\usepackage[T1]{fontenc}
\usepackage{comment}
\usepackage{graphicx,verbatim}

\usepackage[colorlinks=true, urlcolor=blue, citecolor=blue, linkcolor=blue]{hyperref}
\usepackage{color}

\usepackage{amsmath} 
\begin{document}
\title{Automated Optical Density Normalization for Myelin Quantification: Cross-Modal Validation with 7T Ex Vivo MRI}
\titlerunning{Optical Density Normalization for Myelin}
%

\author{Zahra Khodakarami, Sheina Emrani, Pulkit Khandelwal, Chinmayee Athalye, Amanda Denning, Winifred Trotman, Lisa M Levorse, Eric Teunissen-Bermeo, Hamsanandini Radhakrishnan, Daniel Ohm, Christophe Olm, Noah Capp, Ranjit Ittyerah, Karthik Prabhakaran, John A. Detre, Sandhitsu R. Das, David A. Wolk, Corey T McMillan, Gabor Mizsei ,M. Dylan Tisdall, David J Irwin, John L. Robinson, Edward B Lee, Paul A. Yushkevich}  
\authorrunning{Khodakarami et al.}
\institute{University of Pennsylvenia \\
    \email{zkm@seas.upenn.edu}}
  
\maketitle              
\begin{abstract}
White matter hyperintensities (WMH) are bright regions on T2-weighted magnetic resonance imaging (MRI) scans and are associated with cerebrovascular pathology and neurodegeneration, including myelin loss. While Luxol Fast Blue histopathology provides visualization of myelin integrity, quantitative analysis requires measuring Optical Density as a proxy for myelin concentration. However, differences in laboratory protocols and tissue processing introduce staining variability that acts as systematic noise, obscuring the biological signal and preventing consistent comparison across histology runs.
To address this, we developed an automated pipeline that identifies reference (non-pathologic) regions in whole-slide images to compute normalized Optical Density heatmaps. We validated this approach through two complementary evaluations: (1) comparison against expert ratings of myelin loss severity, and (2) cross-modal spatial comparison with co-registered 7T ex vivo MRI for voxel-wise evaluation within white matter regions. The pipeline's reference selection showed strong concordance with expert-identified reference regions, and normalized Optical Density demonstrated a substantially stronger correlation with MRI signal intensity than raw measurements. This correlation persisted within WMH, confirming that the pipeline captures continuous myelin pathology rather than merely the presence or absence of myelin loss contrast. By mitigating staining artifacts, this pipeline provides a robust, validated framework for quantitative cross-modal comparison, establishing a critical methodological foundation for future translation to in vivo myelin mapping and biomarker discovery. Code is available at \href{https://github.com/zkhodakarami/Whole_slide_images}{Github}.

\keywords{Myelin Quantification  \and WMH \and Optical Density}

\end{abstract}

\section{Introduction}
White matter hyperintensities (WMH), regions of hyperintense signal on T2-weighted (T2w) magnetic resonance imaging (MRI), are prevalent neuroimaging markers of aging associated with stroke, dementia, and cognitive decline~\cite{wardlaw2015a}. 
As WMH represent a heterogeneous range of underlying pathologies, including arteriosclerosis, cerebral amyloid angiopathy, and demyelination, moving beyond volumetric measures to pathologically specific biomarkers is critical~\cite{wardlaw2015a,gouw2011a}. However, establishing spatial correspondence between histology and MRI is hindered by the technical variability of histochemical methods. Luxol Fast Blue (LFB) staining is one of the most widely used methods for visualizing myelin sheaths in brain tissue~\cite{inbook,kluver1953}, providing a clear contrast between grey and white matter structures and making it particularly well-suited for cross-modal alignment with MRI. While semi-quantitative ordinal ratings of myelin "pallor" have been the traditional approach, they are inherently subjective, time-consuming, and too coarse to capture subtle variations in myelin density~\cite{deramecourt2012b}. Optical density (OD) measurement offers an objective alternative, capturing stain absorption as a continuous proxy for myelin concentration~\cite{khodanovich2017,warntjes2017,ihara2010a}. However, it remains confounded by variability in tissue preparation and batch effects, which introduce intensity differences across slides unrelated to the underlying biology.

While several computational stain normalization methods, both classical ~\cite{macenko2009,reinhard2001,vahadane2016} 
and deep learning-based~\cite{kang2021}, exist, they standardize color appearance 
relative to external reference images to improve the downstream 
classification tasks. However, these approaches rely on global color 
statistics without encoding what constitutes biologically intact 
myelin, and recent studies have shown that substantial inter-batch 
and inter-laboratory variation persists even after 
normalization~\cite{khan2025,dunn2025,lin2025}. \textbf{To the best of our knowledge, within histopathology stain normalization, there is no method that normalizes stain absorption to the most intact/stable region of each tissue or batch.}

Without a stable reference baseline, OD values make biological interpretation of stain absorption more challenging and hinder comparison across stained slides. \textbf{We reframe stain normalization as a biological reference-discovery problem}: rather than standardizing global color statistics, we automatically identify internal tissue regions where myelin is most likely to be intact and calibrate OD against them, mirroring how neuropathologists implicitly assess demyelination by judging pallor relative to normal-appearing areas within the same section.

A major rate-limiting and time-consuming step in adjusting for staining variability is the requirement for manual annotations. We address this challenge by leveraging \textbf{an inverted-attention mechanism} derived from a CLAM model previously trained to identify histopathological features of cognitive impairment~\cite{mckenzie2022c}. Specifically, we used the model’s ability to detect white matter pallor, a key indicator of myelin loss, to automatically isolate stable reference regions for normalization without manual oversight. By designating areas with the lowest attention scores as candidates for intact myelin, \textbf{we effectively converted a pathology-detection network into an annotation-free reference-discovery algorithm}. Combined with blind stain deconvolution, which requires no predefined stain vectors, this approach enables biologically grounded OD calibration that adapts to each slide's specific staining characteristics.

\begin{figure}
        \centering
        \includegraphics[width=\textwidth]{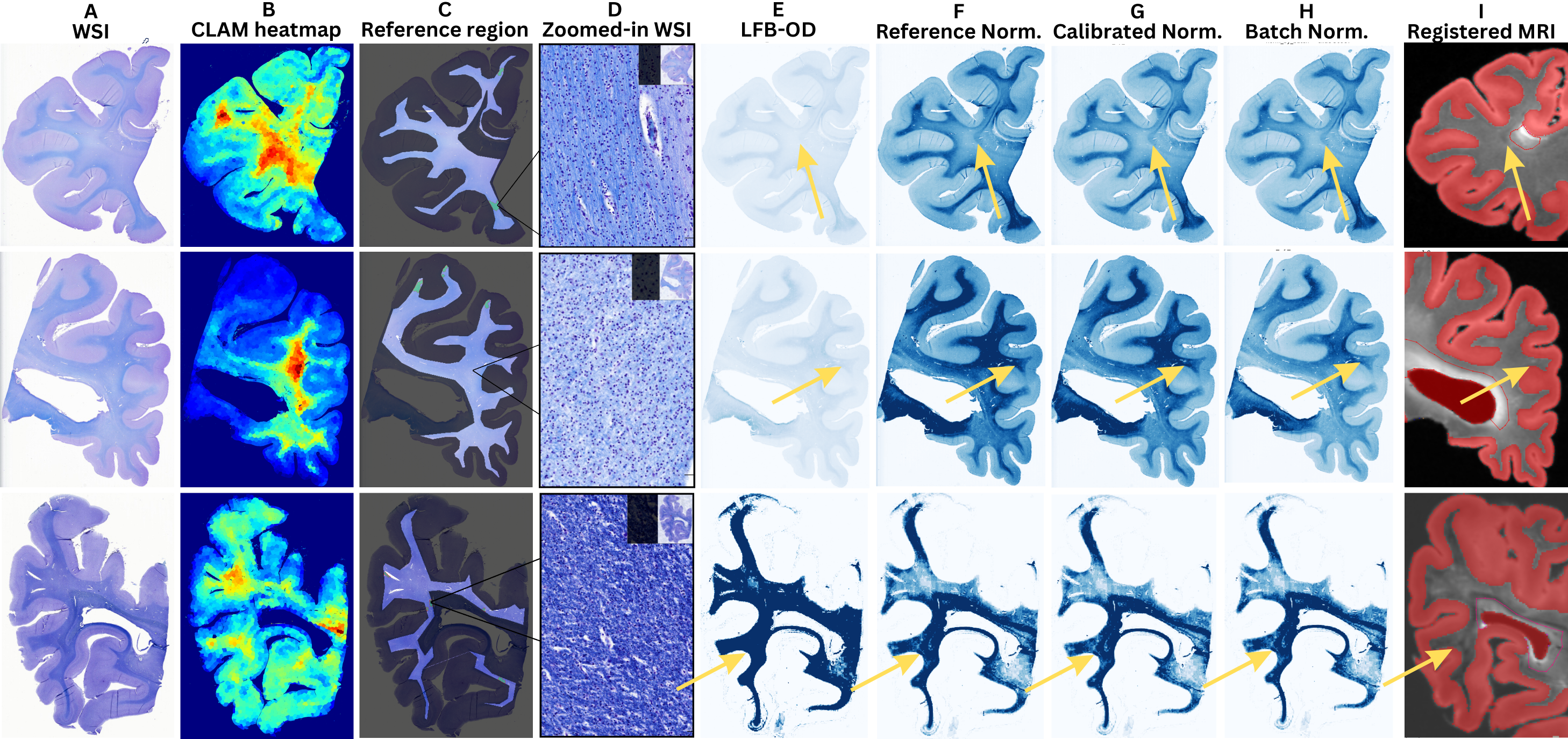}
        \caption{Overview of the proposed pipeline. Details in section 2. The top two rows belong to the same batch, whereas the bottom row is from another hemisphere processed under a different setting. WSI: Whole-Slide Images, LFB-OD: Luxol Fast Blue Optical Density, Norm.: Normalized. (I) white matter regions are not masked red, and delineated areas with lines are white matter hyperintensities; used for cross-modal evaluation(Section~\ref{sec:reg}).}
        \label{fig:pipeline}
\end{figure}

\section{Materials and Methods}
\label{sec:methods}

\textbf{Dataset.} We analyzed 321 LFB-Crystal Violet (CV) stained whole-slide histology 
sections from 45 hemispheres (20F/25M, age $76.22 \pm 10.05$ 
years, PMI $18.76 \pm 9.99$ hours) that underwent ex vivo imaging across the continuum of Alzheimer's disease pathology, imaged at 0.3~mm$^3$ 7T MRI at XYZ center. Human brain specimens were obtained 
with informed consent from next of kin. Following formalin fixation, hemispheres were scanned using a 3D T2 SPACE sequence (0.3~mm$^3$ isotropic, TR/TE = 3000/383~ms, 2--3 hours per scan). 
Tissue slabs were sampled using patient-specific 3D-printed molds, processed into formalin-fixed paraffin-embedded blocks, sectioned between 20-30 microns, and stained with LFB-CV. Whole-slide images were digitized at $0.4 \times 0.4$~$\mu$m$^2$ (20$\times$). A subset of 151 sections from 31 hemispheres (14F/17M, 
age $76.81 \pm 10.61$ years, PMI $17.71 \pm 7.48$ hours) that passed visual quality control for histology--MRI registration was used for cross-modal evaluation.

Our method takes as input LFB-CV-stained whole-slide images of brain tissue and produces normalized optical density (OD) maps of myelin content. Then they are used for cross-modal comparison with registered T2w MRI to compare gold-standard myelin loss heatmaps with WMH on MRI. As illustrated in Fig.~\ref{fig:pipeline}, the pipeline proceeds as follows: given a set of WSI (A), superpixel-level attention heatmaps using the CLAM model were generated (B). The lowest-attention superpixels within the white matter were selected as candidate reference regions (C). Following stain deconvolution to separate LFB and CV channels, OD heatmaps for the LFB channel were generated (E). OD heatmaps were then normalized against the reference region OD values using three strategies of increasing robustness: per-slide, calibrated, and batch-calibrated normalization(F–H). Finally, registered ex vivo T2w MRIs to the histology coordinate spaces were used to evaluate spatial correspondence between normalized myelin-density heatmaps and MRI signal intensity within white matter and WMH regions (I). While the current paper focuses on comparing LFB-derived OD measures to ex vivo MRI, future work will incorporate in vivo FLAIR MRI, which is available in most cases at the center.

\subsection{WSI Processing}
\label{sec:wsi}
\textbf{White Matter Segmentation.} White matter masks were obtained automatically by applying the nnUNet v2~\cite{isensee2021} model to WSIs downsampled to a fixed height of 500px. The model was trained on 256 manually segmented WSI and achieved a mean Dice of 0.896 on 65 held-out test images. The manual segmentation excluded regions that neuropathologists ignore when identifying intact myelin areas: cut edges (darker LFB due to sectioning artifacts) and periventricular tracts 
(disproportionately high OD from dense fiber packing, not representative of typical white matter).

\textbf{Stain Deconvolution and Optical Density.} Per-slide blind 
color deconvolution isolates the LFB channel from the CV counterstain ~\cite{macenko2009,article}. 
Background illumination $I_0$ is estimated from the 99th percentile of 
pixel intensities, and OD is computed via the Beer--Lambert transform: 
$\mathrm{OD} = -\log_{10}(I/I_0)$. Tissue pixels are identified as 
those with $\max_c(\mathrm{OD}_c) \geq 0.05$ and 
$\sum_c \mathrm{OD}_c \leq 3.0$, where $c$ indexes the RGB channels. 
Stain vectors are resolved via SVD of the tissue OD matrix, followed by 
angular peak detection in the two-dimensional projection plane, adapted 
from QuPath's~\cite{bankhead2017a} blind source separation, requiring no predefined 
reference vectors. The OD matrix is then projected onto the estimated 
LFB stain vector to obtain per-pixel LFB optical density, which is 
retained for all downstream analyses.

\textbf{Attention-Based Reference Region Selection.} We repurpose the pretrained CLAM model of McKenzie et al.~\cite{mckenzie2022c} to identify reference white matter for OD normalization. This model was trained to distinguish cognitively impaired from unimpaired donors using hippocampal and frontal cortex tissue, regions highly susceptible to age-related myelin breakdown~\cite{bartzokis2004,bartzokis2007,ihara2010a}. Because myelin pallor is a dominant discriminative feature, the network assigns high attention to white matter with reduced LFB staining. The softmax normalization enforces a complementary constraint: attention on pathology is withdrawn from intact tissue. We exploit this inverse, hypothesizing that the lowest-attention white matter regions will correspond to reference myelin across various diagnostic groups.

Each WSI is tiled at 20$\times$ (256$\times$256\, px, center-cropped to 224$\times$224) for ResNet-50 feature extraction. Patch-level attention scores are softmax-normalized, back-projected to a tile-grid heatmap, and Gaussian-smoothed ($\sigma = 1.5$ tiles). The WSI is independently partitioned into SLIC superpixels~\cite{achanta2012} in CIELAB space ($n{=}2000$, compactness${=}10$, $\sigma{=}1.0$), restricted to those with at least 50\% white matter overlap ($S_{\text{WM}}$). Each superpixel's score $\bar{A}(s)$ is the area-weighted mean of underlying heatmap values. We select as the reference region $S_{\text{ref}} = \{s \in S_{\text{WM}} : \bar{A}(s) \leq \tau_{0.02}\}$, where $\tau_{0.02}$ is the 2\textsuperscript{nd} percentile of attention within $S_{\text{WM}}$, was define the per-slide reference OD as the mean LFB optical density across $S_{\text{ref}}$. Although the CLAM model was trained on a different cohort and scanner, the shared LFB chromogen preserves the relevant feature space; we validated that CLAM-identified references concord with expert-placed intact regions.

\textbf{Normalization Strategies.} We evaluate four strategies in a consistent progression. Let $\mathrm{OD}_\mathrm{ref}$ denote the per-slide CLAM-derived reference and $\mathrm{OD}_\mathrm{max}(b)$ the maximum calibrated reference across all slides in batch $b$ (same subject and staining session). Raw OD is used as-is; internal-normalized divides by $\mathrm{OD}_\mathrm{ref}$; calibrated divides by $1.1 \times \mathrm{OD}_\mathrm{ref}$, where the slope correction derives from the CLAM-vs-expert linear fit (Section~\ref{sec:results}); and batch-normalized divides by $\mathrm{OD}_\mathrm{max}(b)$. All normalized values are capped at 1.5, beyond which transmittance falls below 3\% of incident light ($10^{-1.5}$), exceeding both the reliable dynamic range of brightfield scanning and the optical density of the most densely myelinated white matter in our dataset.

\subsection{MRI-to-Histology Registration.}
\label{sec:reg}
Histology was sampled using an ex vivo 7T T2w MRI-guided protocol~\cite{athalye2025b}, and the MRI was registered to the histology using greedy diffeomorphic registration~\cite{yushkevich2016} with histology as the fixed target. Segmentation of the T2w MRI was 
performed using the purple-mri pipeline~\cite{khandelwal2024a}. N4 bias 
field correction~\cite{tustison2010} was applied to the hemisphere MRI to correct for spatial 
intensity inhomogeneity, using a binary mask derived from the hemisphere
segmentation to restrict correction to intracranial tissue. T2w 
intensities were then z-score normalized using the corpus callosum (CC) as an internal reference, selected because it is the most densely myelinated white matter structure, yielding consistently low T2w 
signal across subjects. On this normalized scale, values near zero 
correspond to intact myelin, while increasingly positive values indicate 
progressively abnormal tissue. The normalized MRI, along with white 
matter masks restricted to normal appearing white matter, WMH, and CC, were warped to histology space via nearest-neighbor interpolation for cross-modality evaluation.

\begin{figure}
    \centering
    \includegraphics[width=0.75\linewidth]{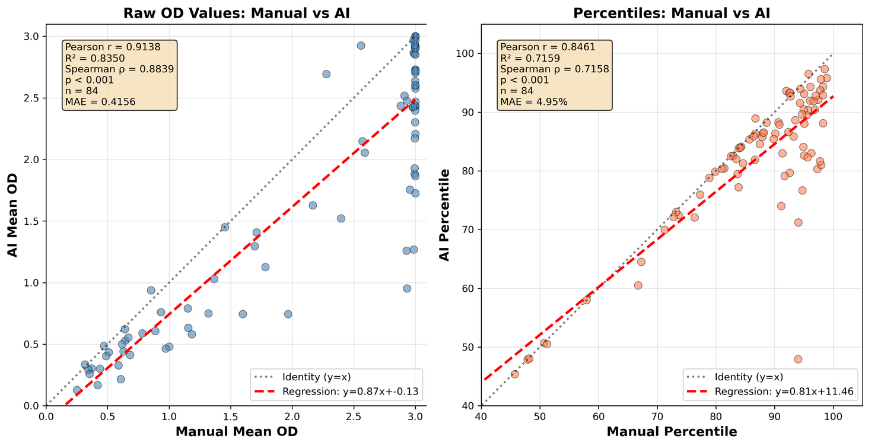}
    \caption{Validation of CLAM-based reference region selection against 
manual expert placement across 84 slides. Left: raw Luxol Fast Blue Optical Density(LFB-OD) values 
(Pearson $r = 0.91$). Right: 
percentile-transformed values (Pearson $r = 0.85$).}
    \label{fig:ref}
\end{figure}

\section{Experiments and Results}
\label{sec:results}
\subsection{WSI Evaluation Results}
\label{sec:refRegion}

To validate CLAM-based reference selection, we compared mean OD of CLAM-identified regions against manually identified intact regions across 84 slides. Two trained raters independently placed reference regions on each slide; discrepant placements were reviewed jointly and resolved by consensus. The consensus reference OD showed strong agreement with CLAM-derived reference OD (Pearson $r = 0.91$, Spearman $\rho = 0.88$, $p < 0.001$). The regression slope of 0.87 motivated the $\beta=1.15$ calibration correction in the caliberated normalization strategy (Fig. \ref{fig:ref}).

\subsection{Expert Pathology Grading}
\label{expert}
\begin{figure}
    \centering
    \includegraphics[width=0.9\textwidth]{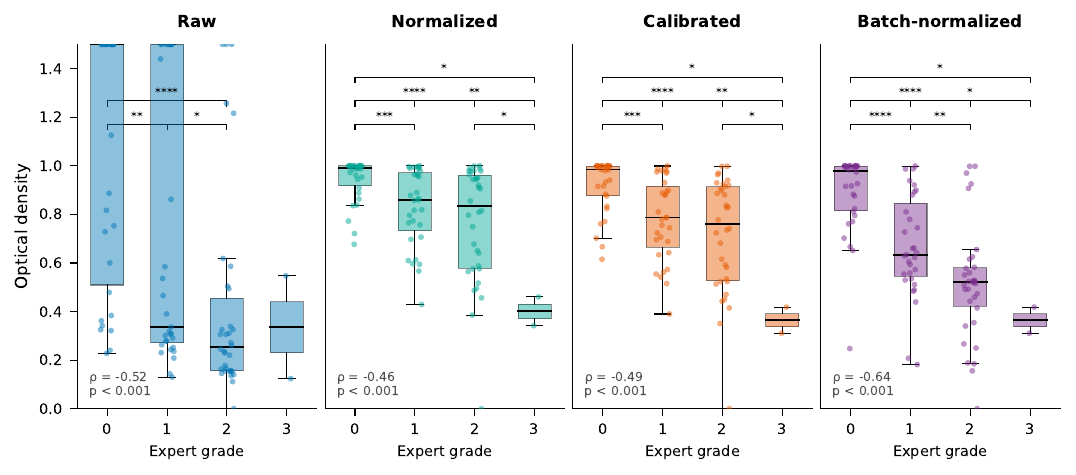}
    \caption{Distribution of Luxol Fast Blue Optical Density by expert pathology 
grade (0~=~intact, 3~=~severe myelin loss; $n = 97$ ROIs) across 
four quantification methods. More details in section~\ref{expert}. Statistical comparisons: 
$*\,p < 0.05$, $**\,p < 0.01$, $***\,p < 0.001$, 
$****\,p < 0.0001$.}
    \label{fig:expert}
\end{figure}

We assessed four OD quantification strategies against expert 
pathology ratings across 97 ROIs rated on the 0--3 myelin 
loss scale of Deramecourt et al. ~\cite{deramecourt2012b}. Two trained raters independently rated each ROI; discrepant scores were reviewed jointly and resolved by consensus. Batch-normalized OD achieved the strongest correlation with 
consensus grades ($\rho = -0.64$, $p < 0.001$), followed by 
raw ($\rho = -0.52$), calibrated ($\rho = -0.49$), and internal-normalized OD ($\rho = -0.46$). Batch normalization also yielded the highest discrimination between intact (grade~0) and demyelinated (grades~2--3) tissue (AUC $= 0.90$), compared with raw (AUC $= 0.85$), calibrated (AUC $= 0.82$), and internal-normalized (AUC $= 0.81$). Raw OD exhibited extreme technical variability among grade-0 
(intact) regions (CV $= 178.6\%$), driven in part by 
non-physical values arising from tissue folds and deconvolution 
artifacts. Even after truncation at OD $= 1.5$, raw CV remained 
high ($49.4\%$), whereas internal normalization reduced it to 
$9.6\%$, demonstrating that reference-based normalization 
addresses both artifactual outliers and genuine staining 
heterogeneity(Fig.~\ref{fig:expert}).

\subsection{Cross-Modal Evaluation Results}

We evaluated voxel-wise correspondence between LFB OD and T2w MRI across 
151 co-registered sections from 31 hemispheres within full white matter 
and WMH masks. Given the CC z-score normalization, 
the expected relationship is negative: high OD reflects intact myelin 
(near-zero z-score), whereas demyelination yields reduced OD and elevated 
T2w signal. Table~\ref{tab:crossmodal} summarizes all metrics.

Batch normalization achieved the strongest voxel-wise Spearman correlation 
($\rho = -0.379 \pm 0.204$), significantly outperforming raw OD, 
internal, and calibrated normalization across all metrics. Given the 
inherent limitations of cross-modal histology--MRI comparison, including 
microscopic registration errors and resolution mismatch, these 
correlations represent robust correspondence~\cite{athalye2025b}. Shift analysis confirmed 
impact of residual registration error (best $\rho$ within 
$\pm$3 pixels improved from $-0.379$ to $-0.424$). Moreover, spatial 
overlap was assessed by independently thresholding each modality (bottom 
25th percentile OD; top 75th percentile T2w) and computing Dice and 
AUC-ROC between the resulting lesion maps. To account for 
non-independence of multiple sections per hemisphere ($4.9 \pm 3.3$), 
paired $t$-tests on hemisphere-level means confirmed all normalization 
methods significantly outperformed raw OD (all $p < 10^{-5}$), with 
batch normalization significantly exceeding internal ($p = 1.0 \times 
10^{-4}$) and calibrated ($p = 6.8 \times 10^{-3}$) normalization. 
Within WMH regions, the correlation persisted ($\rho = -0.336 \pm 0.303$), 
confirming that cross-modal agreement reflects a continuous relationship 
within affected white matter.
\begin{table}
\caption{Cross-modal evaluation of Luxol Fast Blue Optical Density against T2w ex vivo MRI 
across 151 histology sections. \textbf{Values are reported as $\times 10^{-2}$ 
(mean $\pm$ SD)}. Across all normalized methods, 148 of 151 sections 
(98\%) achieved significant voxel-wise correlation ($p < 0.05$) within white matter, and 108 of 125 sections (86\%) within WMH regions. 
Best shift $\rho$: best Spearman correlation achievable within a 
$\pm$3-pixel spatial shift, quantifying sensitivity to residual 
registration error. Best values per metric in 
\textbf{bold}.}\label{tab:crossmodal}
\begin{tabular}{|l|l|l|l|l|}
\hline
Metric & Raw & Internal-norm. & Calibrated & Batch-norm. \\
\hline
\multicolumn{5}{|l|}{\textit{All white matter (n = 151)}} \\
\hline
Spearman $\rho$         & $-23.8 \pm 22.7$ & $-35.6 \pm 19.8$ & $-36.6 \pm 20.0$ & $\mathbf{-37.9 \pm 20.4}$ \\
Best shift $\rho$       & $-27.9 \pm 26.1$ & $-40.3 \pm 22.1$ & $-41.3 \pm 22.3$ & $\mathbf{-42.4 \pm 23.1}$ \\
Pearson $r$             & $-19.6 \pm 21.1$ & $-31.0 \pm 19.6$ & $-32.6 \pm 20.0$ & $\mathbf{-33.5 \pm 19.7}$ \\
Dice coefficient        & $32.9 \pm 17.6$  & $42.6 \pm 14.0$  & $42.8 \pm 14.1$  & $\mathbf{43.2 \pm 14.0}$  \\
Area under ROC          & $62.0 \pm 12.1$  & $68.8 \pm 11.3$  & $69.3 \pm 11.5$  & $\mathbf{69.8 \pm 11.6}$  \\
\hline
\multicolumn{5}{|l|}{\textit{WMH only (n = 125)}} \\
\hline
Spearman $\rho$         & $-23.0 \pm 28.1$ & $-32.8 \pm 29.8$ & $-33.1 \pm 29.9$ & $\mathbf{-33.6 \pm 30.3}$ \\
Pearson $r$             & $-22.8 \pm 26.8$ & $-33.0 \pm 28.4$ & $-33.4 \pm 28.3$ & $\mathbf{-33.4 \pm 28.2}$ \\
\hline
\end{tabular}
\end{table}
\section{Discussion}

LFB staining intensity varies substantially across slides due to 
fixation, section thickness, and staining protocol differences, 
producing a continuous OD gradient whose absolute values are not 
directly comparable across sections. Our attention-based approach 
addresses this by automatically identifying intact white matter as an 
internal reference ($r = 0.91$ with expert selection), while batch 
normalization further removes inter-session variability, reducing 
the coefficient of variation within intact regions from 178.6\% to 
18.5\%. Furthermore, the pipeline is robust to domain shifts in tissue origin and scanner type. Because the shared LFB chromogen preserves the underlying feature space of myelin pallor, the repurposed CLAM model generalizes well beyond its original training data in the hippocampus and frontal cortex.

To our knowledge, the cross-modal validation against the co-registered 
7T MRI across 151 sections from 31 hemispheres represents the largest 
voxel-wise histology--MRI correlation study of myelin integrity at 
this resolution. The observed correlations ($\rho = -0.379$) are 
notable given inherent cross-modal limitations: registration errors, 
resolution mismatch, and the distinct physical bases of LFB OD and T2 
relaxation. Indeed, as demonstrated by our spatial shift analysis, even microscopic misalignments severely artificially suppress the theoretical maximum correlation. That the correlation persists within WMH regions 
($\rho = -0.336$) confirms the pipeline captures a continuous spectrum 
of myelin pathology, not merely lesional versus non-lesional contrast. This quantitative link between histological myelin density and MRI 
signal opens a concrete downstream application: using the 
histology-derived OD maps of myelin as voxel-level ground truth to train or 
calibrate models that predict regional myelin loss from in vivo FLAIR 
MRI.

Several limitations should be noted. Analysis were conducted with a single MRI 
protocol; generalizability to other sequences remains to be 
established. Batch normalization requires multiple slides from the 
same staining session, though internal normalization provides a 
practical single-slide alternative.

\textbf{Conclusion.} We presented a novel reference-discovery algorithm that leverages an inverted-attention mechanism and blind stain deconvolution for automated, quantitative myelin assessment in LFB-CV histology. Validated against expert ordinal ratings and through the largest voxel-wise histology-MRI correlation study of its kind (151 co-registered 7T MRI and histology sections), our results demonstrate that batch-calibrated normalization yields the most consistent and biologically grounded measure of myelin integrity. By successfully mitigating systemic staining variability, this pipeline captures myelin loss and establishes a robust histochemical foundation for the development and calibration of in vivo imaging biomarkers.

%
%
%
\bibliographystyle{splncs04}
\bibliography{Miccai2026}

\end{document}